\documentclass[prl,showpacs,twocolumn]{revtex4}
\usepackage{graphicx}
\usepackage{bm}

\def\zexpo{\theta}
\def\avereps{\overline{\varepsilon}}
\def\sectionskip{\medskip}

\begin{document}

\title{Glass transition in secondary structures formed
by random RNA sequences}
\author{R.~Bundschuh}
\author{T.~Hwa}
\affiliation{
Department of Physics, University of California at San Diego,
La Jolla, CA  92093-0319, U.S.A.}
\affiliation{Institute for Theoretical Physics, University of California,
Santa Barbara, CA 93106-4030, U.S.A.}
\date{\today}

\begin{abstract}
Formation of RNA secondary structures is an example of the sequence-structure
problem omnipresent in biopolymers. A theoretical question of recent interest is
whether a random RNA sequence undergoes a finite temperature glass transition.
We answer this question affirmatively by first establishing the perturbative
stability of the high temperature phase via a two replica calculation.
Subsequently, we show that this phase cannot persist down to zero temperature
by considering energetic contributions due to rare regions of 
complementary subsequences.

\end{abstract}

\maketitle

\noindent
{\it Introduction:} RNA is an important biopolymer critical to all
living systems~\cite{rna}. Like in DNA, there are four types of
nucleotides (or bases) A, C, G, and U which, when polymerized can form
double helical structures consisting of stacks of stable Watson-Crick
(A--U or G--C) pairs.  However unlike a long polymer of DNA, which is
often accompanied by a complementary strand and forms otherwise
featureless double helical structures, a polymer of RNA is usually
single-stranded. It bends onto itself and forms elaborate spatial
structures in order for bases located on different parts of the
backbone to pair with each other. The structures encoded by the
primary sequences often have important biological functions, much like
how the primary sequence of a protein encodes its structure.

Understanding the encoding of structure from the primary sequence has
been an outstanding problem of theoretical biophysics. Most work in
the past decade has been focused on the problem of protein folding,
which is very difficult analytically and numerically~\cite{protein}.
Here, we study the problem of RNA folding, specifically the formation
of RNA {\em secondary structures} which is more amenable to analytical
and numerical approaches due to a separation of energy
scales~\cite{tino99}. Efficient algorithms~\cite{mcca90,rnaalg}
together with carefully measured free energy parameters~\cite{frei86}
describing the formation of various microscopic structures (e.g.,
stacks, loops, hairpins, etc.) allow the exact calculation of the
ensemble of secondary structures formed by a given RNA molecule of up
to a few thousand bases.

In this work, we are not concerned with the structure formed by a
specific sequence. Instead, we will study the statistics of secondary
structures formed by the ensemble of {\em long} 
{\em random} RNA sequences. It has been
debated recently whether such an ensemble undergoes a finite temperature
glass transition~\cite{higg96,pagnani}. 
However, the numerical results these studies are based on are not clear
enough to allow unambiguous interpretation.
In this letter, we provide {\em analytical evidence} supporting
the existence of a finite temperature glass transition by studying some
toy models of RNA folding. We characterize the
behavior of RNA in the absence of disorder and show with the help
of a two-replica calculation that disorder is perturbatively irrelevant. 
We then show that the assumption that the high-temperature
phase exists at all temperatures leads to a contradiction below
some finite temperature, thereby necessitating a finite temperature phase
transition.

\begin{figure}[hbtp]
\begin{center}
\vspace*{-6mm}
\includegraphics[angle=0,width=0.9\columnwidth]{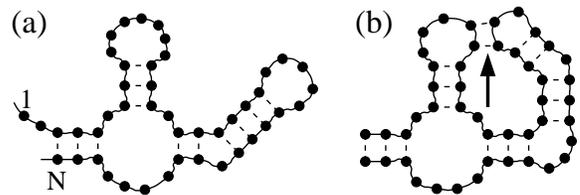}
\vspace*{-5mm}
\caption{Secondary structures of an RNA molecule: The solid and dashed
lines represent the backbone and the base pairs respectively. (a)
shows a valid secondary structure while (b) contains a pseudo-knot as
indicated by the arrow.}
\label{fig_strdiags}
\vspace*{-8mm}
\end{center}
\end{figure}

\sectionskip
\noindent
{\it Model:} A secondary structure $S$ of a polymer of RNA is the set
of pairings formed between all of its monomers (or bases), with each
base allowed in at most one pairing. We denote the pairing
between the $i^{\rm th}$ and $j^{\rm th}$ monomer
by $(i,j)$, with $1\!\le\!i\!<\!j\!\le\!N$. Each such
structure can be represented by a diagram like the one shown in
Fig.~\ref{fig_strdiags}(a).  In addition, it is common to exclude
``pseudo-knots'' like the one shown in Fig.~\ref{fig_strdiags}(b) from
the definition of secondary structures, so that any two base pairs
$(i,j)$ and $(k,l)$ are either independent, i.e.,
$i\!<\!j\!<\!k\!<\!l$, or nested, i.e., $i\!<\!k\!<\!l\!<\!j$.  This
is permissible since long pseudo-knots are kinetically difficult to
form and even the short ones occur infrequently due to energetic
reasons~\cite{tino99}. Experimentally, it is possible to ``turn off''
the pseudo-knots and other complicated tertiary contacts~\cite{tino99}
and study exclusively the class of secondary structures defined above.

In order to calculate Boltzmann
factors for an ensemble of secondary structures, we need to assign
an energy $E[S]$ to each structure $S$. For the purpose of secondary
structure {\em prediction}, it is essential to model the energy
as accurately as possible~\cite{mcca90,rnaalg}.
However, for our interest in the universal statistical properties of long,
random RNA sequences far below the denaturation temperature, it suffices
to consider simplified models along the line of those used in earlier 
studies~\cite{higg96,pagnani,bund99,higg00}.

We associate an interaction energy $\varepsilon_{i,j}$ 
with every pairing $(i,j)$ and assign $E[S]=\sum_{(i,j)\in S}\varepsilon_{i,j}$
as the total energy of the structure $S$. To retain the spirit of
Watson-Crick pairing, we choose random sequences $b_1\ldots b_N$ of the
four bases $A$, $U$, $C$, and $G$ and then assign
\begin{equation}\label{eq_energychoice}
\varepsilon_{i,j}=\left\{\begin{array}{rl}-u_m&
\mbox{$(b_i,b_j)$ is a Watson-Crick base pair}\\
u_{mm}&\mbox{otherwise}\end{array}\right.
\end{equation}
with $u_m,u_{mm}\!>\!0$.  Here, $u_m$ sets the energy scale.  The
value of $u_{mm}$ is not essential as long as it is repulsive, since
there is always the option to not bind at all (with energy
``0'') rather than to bind with a repulsive energy.

For analytical work, it is convenient to take the
$\varepsilon_{i,j}$ to be {\em independent Gaussian} random variables
specified by the mean $\avereps$ and variance
\begin{equation}\label{eq_gaussianmoments}
\overline{(\varepsilon_{i,j}-\avereps)(\varepsilon_{k,l}-\avereps)}
=D\,\delta_{i,k}\delta_{j,l}.
\end{equation}
Throughout the text, we will use the overline to denote averages over the 
ensemble of random pairing energies.
Here, the parameter $D$ provides a measure of the strength of the randomness.
It is an approximation to the model (\ref{eq_energychoice}) in two respects:
First, it replaces the
discrete distribution of energies by a Gaussian
distribution. Moreover, it neglects the {\em correlations} between
$\varepsilon_{i,j}$ and $\varepsilon_{i,k}$ induced by the shared base
$b_i$. We do not anticipate universal quantities to depend on the
subtle differences in the statistics of the
$\varepsilon_{i,j}$'s. This will be tested numerically by comparing
the scaling behavior produced by the two models.

Given the energy of each secondary structure, we can study the
partition function $Z(N)\!=\!\sum_{S\in{\cal S}(N)}e^{-\beta E[S]}$
where ${\cal S}(N)$ comprises all valid
secondary structures of a molecule of length $N$.
This function can be conveniently computed in
terms of the {\em restricted} partition function $Z_{i,j}$ for the
substrand $b_i \ldots b_j$. For the simple energy model $E[S]$
described above, $Z_{i,j}$ can be split up according to the possible
pairings of position $j$ to yield the exact recursion
relation~\cite{higg96,dege68,wate78}
\begin{equation}
Z_{i,j}=Z_{i,j-1} +
\sum_{k=i}^{j-1} Z_{i,k-1}\cdot e^{-\beta\varepsilon_{k,j}} \cdot
Z_{k+1,j-1}.\label{eq_partfuncrec}
\end{equation}
This equation can then be iterated to compute the full partition function
$Z(N)\!=\!Z_{1,N}$ for {\em arbitrary} interaction energies 
$\varepsilon_{i,j}$'s in $O(N^3)$
time~\cite{mcca90,wate78}. It
also forms the basis of analytical approaches to the problem.

\sectionskip
\noindent{\it The molten phase:} If sequence disorder does not play an
important role, we may describe the RNA molecule by replacing all the
binding energies $\varepsilon_{i,j}$ by some effective value
$\varepsilon_0$. As we will see later, this is an adequate
description of random RNA at high enough temperatures (but below
denaturation.)  Below we briefly review the properties of RNA
in this high temperature ``molten phase''.

Since the $Z_{i,j}$'s become 
translationally invariant in the absence of disorder,
it is straightforward to solve Eq.~(\ref{eq_partfuncrec}) for the molten 
phase partition function $Z_0(N)$ using the $z$-transform. For large $N$,
it has the form~\cite{bund99,dege68,wate78}
\begin{equation}\label{eq_moltenz}
Z_0(N)\stackrel{N\gg 1}{\sim} N^{-\zexpo_0}\exp[-\beta N f_0(q)]
\end{equation}
where  $q\!\equiv\!e^{-\beta\varepsilon_0}$ is the only parameter, 
$\zexpo_0\!=\!3/2$ is a universal exponent, and 
$f_0(q)\!=\!-k_B T\ln(1\!+\!2\sqrt{q})$ is the free energy per length. 

One useful observable characterizing the state of the RNA is 
the free energy cost $\Delta F(N)$ of {\em pinching together} 
one end (say $i=1$) and the mid-point ($i=N/2$) of the polymer 
relative to the unperturbed state. The pinch effectively separates the
polymer into two pieces of length $N/2$. In the molten phase, 
we simply have
\begin{equation}\label{eq_moltenpinch}
\Delta F = -k_BT \ln [Z_0(N)/Z_0^2(N/2)] \approx
{\textstyle\frac{3}{2}}k_B T \ln N.
\end{equation}
It  reflects the loss of {\em configurational entropy} in the allowed
secondary structures due to the pinch.

\begin{figure}[hbtp]
\begin{center}
\vspace*{1mm}
\includegraphics[angle=0,width=0.9\columnwidth]{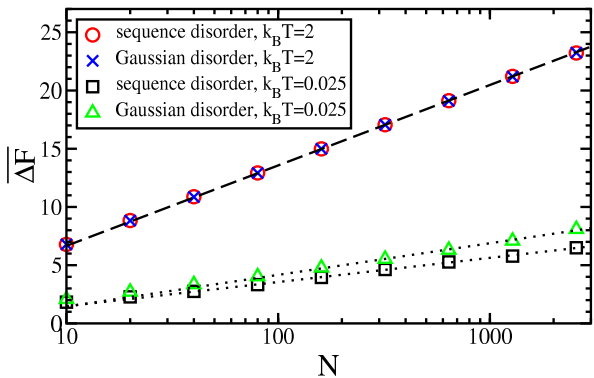}
\vspace*{-3mm}
\caption{Ensemble averaged pinching free energy $\overline{\Delta F}(N)$
for two different choices of
disorder energies according to Eqs.~(\protect\ref{eq_energychoice})
and~(\protect\ref{eq_gaussianmoments}), at \protect$k_BT=2$ and
\protect$k_BT=0.025$. The statistical error is no larger 
than the size of the symbols. At \protect$k_B T=2$, the data follows
precisely the expected $3\ln N$ behavior (dashed
line.) At \protect$k_BT=0.025$, the best fits for the data at
\protect$N\ge160$ to a logarithmic behavior (dotted lines) give
\protect$\overline{\Delta F}\approx0.90\ln N$ for sequence and
\protect$\overline{\Delta F}\approx1.18\ln N$ for Gaussian disorder.}
\label{fig_pinch}
\vspace*{-8mm}
\end{center}
\end{figure}

\sectionskip
\noindent{\it Numerics:} Before we delve into the analysis,  
we first present some numerical evidence for a phase
transition between the high- and low- temperature behavior
for the ensemble of random RNA. To this end, we generate many
configurations of interaction energies $\varepsilon_{i,j}$'s
according to both models (\ref{eq_energychoice})
and~(\ref{eq_gaussianmoments}), with $u_m\!=\!u_{mm}\!=\!1$ and
$\avereps\!=\!1/2$, $D\!=\!3/4$ 
respectively.  Then for a wide range of temperatures
we calculate the pinching free energy, which is given by
$\Delta F\!=\!-k_{\mathrm{B}}T
\ln(e^{-\beta\varepsilon_{1,N/2}}Z_{2,N/2-1}Z_{N/2+1,N}/Z_{1,N})$
in terms of the quantity $Z_{i,j}$ in (\ref{eq_partfuncrec})
for sequences without translational invariance.
The result is then averaged 
over $1000$ to $10000$ disorder configurations and illustrated in
Fig.~\ref{fig_pinch} for two representative temperatures.

At high temperature ($k_B T\!=\!2$), the pinch energy follows precisely the molten phase
behavior expected according to Eq.~(\ref{eq_moltenpinch}) up to an additive
constant. Thus, at
high temperatures the molten phase description is applicable even if
the interaction energies $\varepsilon_{i,j}$ are not
uniform. Moreover, we find no difference between the two models of 
disorder at $k_B T\!=\!2$. At low temperature ($k_BT=0.025$), the picture is
different. The length dependence of $\Delta F$
is still logarithmic. However, the prefactors differ between
the two disorder choices and they are both by far larger than the
prefactor $0.0375$ expected if the molten phase result is extrapolated
to this temperature. This suggests that
there is a distinct low temperature phase;
this finding is reinforced by similar changes detected in other 
observables~\cite{bund01}.

\sectionskip
\noindent
{\it High temperature behavior:} Now we will establish the stability
of the molten phase against weak disorder by a perturbative analysis.
Assuming that the specific choice of disorder does not
matter at weak disorder (or high temperature) as supported by the
numerical results above, we will use the uncorrelated Gaussian disorder
characterized by Eq.~(\ref{eq_gaussianmoments}) for the purpose of this
analysis. The behavior of the system at weak disorder is determined by 
the lowest order terms in the perturbative expansion
of the ensemble averaged free energy in the disorder strength
$D$.  This term is given by the two-replica partition function
$\overline{Z^2(N)}$ of two RNA molecules sharing the same disorder.
With the uncorrelated Gaussian energies (\ref{eq_gaussianmoments}), the
ensemble average can be explicitly performed, yielding after some
algebra
\begin{equation}\label{eq_tworpartfunc}
\overline{Z^2(N)}=\sum\nolimits_{\{S_1,S_2\in{\cal S}(N)\}}q^{|S_1|}q^{|S_2|}
\widetilde{q}^{|S_1\cap S_2|}
\end{equation}
where $q\equiv\exp\left(-\beta\avereps+\frac{1}{2}\beta^2D\right)$ and
$\widetilde{q}\equiv\exp\left(\beta^2D\right)$ are the two relevant
``Boltzmann factors'', and $|S_i|$ and $|S_i\cap S_j|$ are the number
of bases contained in structure $S_i$ or common to $S_i$ and $S_j$ respectively.
This effective partition function has a simple physical
interpretation: It describes two RNA molecules subject to a {\em
homogeneous} attraction with effective interaction energy
$\varepsilon_0\!\equiv\!-\beta^{-1}\ln q\!=\!\avereps\!-\!\frac{1}{2}\beta D$
between any two bases of the same molecule. In addition, there is an
inter-replica attraction characterized by the factor $\widetilde{q}$
for each bond {\em shared} between the two replicas. The inter-replica
attraction is induced by the same disorder shared by the
replicas. It can potentially force the replicas to ``lock'' together,
i.e., to become correlated.

It turns out that this two-replica problem can be solved
exactly~\cite{bund01}.  The key idea leading to the solution is that the sum
over the pairs of secondary structures in Eq.~(\ref{eq_tworpartfunc})
can be reordered by first summing over the bonds which are {\em
common} to the two structures, noting that the possible configurations of the
common bonds are themselves the set ${\cal S}(N)$ of valid secondary
structures. Within a given
configuration of common bonds, all possibilities to place non-common
bonds in the two individual structures can then be summed over, leading
to an effective single RNA problem. However, the necessary
algebra is quite involved~\cite{bund01}; here we just quote the
results. The solution has the form $\overline{Z^2(N)}\sim N^{-\zexpo}
\exp[-\beta N f(q,\widetilde{q})]$, with two different expressions for 
$\zexpo$ and $f$ depending on whether $\widetilde{q}$ is above or below a
critical value $\widetilde{q}_c = 1+ 1/[q^2 \sum_{N=1}^\infty N g^2(N)]$,
where $g(N) = Z_0(N)/(1\!+\!2\sqrt{q})^{N-1}\!\sim\!N^{-3/2}$ for large $N$.

For $\widetilde{q}\!<\!\widetilde{q}_c$, we have
$\zexpo\!=\!2\zexpo_0\!=\!3$. In addition, $f(q,\widetilde{q})$ is
basically a modified version of $2f_0(q)$. In this regime, the
two-replica partition function $\overline{Z^2(N)}$ is essentially a
product of two single-replica partition functions $Z_0(N)$. Since there is
no coupling between the two replicas, we conclude that the effect of disorder 
is irrelevant in this regime.
For $\widetilde{q} > \widetilde{q}_c$, we have $\zexpo = 3/2$ and $f$
given by a complicated function of $Z_0(N)$~\cite{bund01}.
Here, the two-replica partition function
has the {\em same} asymptotic form as that of the
single-replica system in (\ref{eq_moltenz}).
This implies that the disorder coupling {\em locks} the two replicas together.

Of course, as already explained above, only the weak-disorder limit
(i.e., $\widetilde{q}\gtrsim\!1$) of the two-replica problem can be
applied to the full random RNA problem. While $\widetilde{q}_c$ itself
depends on the disorder strength $\beta^2 D$, it converges in the weak
disorder limit ($\beta^2D\ll 1$) towards a constant
$\widetilde{q}_c(\beta^2D\!=\!0)\!>\!1$. Therefore, for $\beta^2 D\ll1$ we
always have $\widetilde{q}<\widetilde{q}_c$. The molten phase is
perturbatively stable upon the introduction of disorder, making it the
appropriate description of random RNA at high temperatures.

\sectionskip
\noindent
{\it Low temperature behavior:} Next we determine whether the molten
phase persists for all temperatures down to $T=0^+$.
In the following,  we will {\em assume} that long
random RNA is in the molten phase for {\em all} temperatures, i.e., that
the partition function for any substrand of large length is
given by Eq.~(\ref{eq_moltenz}), with some effective value of
$q$. Then, we will show that this assumption leads to a contradiction
below some temperature $T^*>0$. This contradiction implies 
that the molten phase description breaks
down at some finite $T_c \ge T^*$. To be specific, we will consider the
sequence disorder model (\ref{eq_energychoice}) in this analysis.

We will again focus on the pinching free energy $\Delta F$.
Under the assumption that the random sequences are described by the
molten phase, it is given by Eq.~(\ref{eq_moltenpinch}) for large $N$
and all $T$ {\em independently} of the effective value of $q$. On the
other hand, we can study this pinching free energy for each given
sequence of bases drawn from the ensemble of random sequences. 
For each such sequence, we can look for a
continuous segment of $\ell\ll N$ Watson-Crick pairs
$(b_i,b_j)(b_{i+1},b_{j-1})\ldots(b_{i+\ell-1},b_{j-\ell+1})$ where
the bases $b_i\ldots b_{i+\ell-1}$ are within the first half of the
molecule and the bases $b_{j-\ell+1}\ldots b_j$ are in the second
half. For random sequences, the probability of finding such complementary segments
decreases exponentially with the length $\ell$ in a given region,
with the largest $\ell$ in a sequence of length $N$ being typically 
$\ln N/\ln2$~\cite{arra90}.

Now we calculate the pinching free energy $\Delta
F=F_{\mathrm{pinched}}-F_{\mathrm{unpinched}}$ by evaluating the two
terms separately. The partition function corresponding to the
unpinched free energy contains {\em at least} all the configurations
in which the two complementary segments $b_i\ldots b_{i+\ell-1}$ and
$b_{j-\ell+1}\ldots b_j$ are completely paired.  Thus,
\begin{equation}\label{eq_unpinchedandpaired}
F_{\mathrm{unpinched}}\le F_{\mathrm{paired}}
\end{equation}
where $F_{\mathrm{paired}}$ is the free energy of the ensemble of
structures in which the two complementary segments are paired. The latter
 is the sum of the energy of the paired segments and those
of the two remaining substrands 
of lengths $L_1\!=\!j\!-\!i\!-\!2\ell\!+\!1$ and 
$L_2\!=\!N\!+\!i\!-\!j\!-\!1$, i.e.,
\begin{equation}\label{eq_fpaired}
F_{\mathrm{paired}}\!\!=\!\!-\ell u_m \!+ (N\!-\!2\ell) f_0
+{\textstyle\frac{3}{2}}k_B T\big[\!\ln(L_1)\!
+\!\ln(L_2)\big].
\end{equation}

The free energy $F_{\mathrm{pinched}}$ is, by the assumption of
the molten phase, the interaction energy of the pinched base pair $(b_1,b_{N/2})$ 
plus the free energy of the two substrands $b_2\ldots b_{N/2-1}$ and 
 $b_{N/2+1}\ldots b_N$. According to Eq.~(\ref{eq_moltenz}), this is
$F_{\mathrm{pinched}}=f_0N+2\times\frac{3}{2}k_B T\ln N$ up to terms
independent of $N$.  Combining this with
Eqs.~(\ref{eq_unpinchedandpaired}) and~(\ref{eq_fpaired}), and noting that
$\ell$ is typically of the order $\ln N/\ln 2$ and $L_1$, $L_2$ are typically
proportional to $N$, we finally obtain $\overline{\Delta F}\ge
[u_m+2f_0]\ln N/\ln2$ for very large $N$. This is only consistent with
Eq.~(\ref{eq_moltenpinch}) if
\begin{equation}\label{eq_consistencycondition}
{\textstyle\frac{3}{2}}k_B T\ge [u_m+2f_0]/\ln2.
\end{equation}

Since $f_0$ is a free energy per length, the right hand side of this
equation is a decreasing function of temperature.  Moreover,
$f_0(T\!=\!0)$ is the average free energy per length of the minimum
energy secondary structures of random sequences. The lowest possible
energy any structure of an RNA molecule of $N$ bases can achieve is
$-\frac{N}{2}u_m$, i.e.,  if {\em all} bases form Watson-Crick 
pairs. However, if the sequence disorder leads to frustration,
there is always a finite fraction of bases which cannot be
incorporated into Watson-Crick base pairs~\footnote{This is not the
case for the sequence disorder model at an alphabet size $2$ unless a
minimum hairpin length constraint is introduced as in
Ref.~\cite{pagnani}.} and thus $f_0(T\!=\!0)\!>\!-u_m/2$.  Therefore, the
right hand side of (\ref{eq_consistencycondition}) is a decreasing
function of $T$ starting at some positive value at $T=0$.  It follows that
there is some unique temperature $T^*$ below which the consistency
condition~(\ref{eq_consistencycondition}) breaks down, implying the
inconsistency of the molten phase assumption in this regime.
From this we conclude that there must be a phase transition away from 
the molten phase at some critical temperature $T_c\!\ge\!T^*\!>\!0$.
Numerically, we find that $f_0(T\!=\!0)\!\approx\!-0.46$ for $u_m\!=\!1$,
yielding $k_B T^*\!\approx\!0.075$.
This is consistent with the numerically observed change
between the low- and high- temperature behaviors at 
$k_B T_c\!\approx\!0.25$~\cite{bund01}. Improved estimates of $T_c$ can be 
made by relaxing the condition of perfect complementarity
of the two segments; this as well as a detailed characterization 
of the low temperature phase will be discussed elsewhere.

\sectionskip
\noindent
{\it Conclusions:} We studied the behavior of random RNA sequences in
the regime below the denaturation transition. A two-replica
calculation shows that disorder is perturbatively irrelevant,
i.e., an RNA molecule with weak sequence disorder is in the molten phase
where many secondary structures with comparable total energy
coexist. By further considering the rare regions of strong sequence
complementarity, we show that the molten phase cannot exist down to arbitrarily
low temperatures, and thereby 
establish the existence of a distinct low temperature phase below
some critical temperature $T_c>0$.
Our analysis follows the approach used in Ref.~\cite{tang00} to show the
relevance of disorder in the denaturation of double-stranded DNA.
It will be interesting to see whether a renormalization group theory
along the line of Ref.~\cite{tang00} may be constructed to elucidate the
low-temperature phase of the random RNA problem.

\noindent
{\it Acknowledgments:} The authors benefited from discussions with
U.~Gerland, D.~Moroz, and especially L.-H. Tang regarding the role
of rare regions. This work is supported by the
NSF through grants no.\ DMR-9971456 and DBI-9970199.


\begin{thebibliography}{99}
%
\bibitem{rna}
see, e.g., R.F. Gesteland and J.F. Atkins \textit{eds.}, \textit{The RNA
world : the nature of modern RNA suggests a prebiotic RNA world}
(Cold Spring Harbor Laboratory Press, Cold Spring Harbor, 1993). 
%
\bibitem{protein}
K.A. Dill \textit{et al.}, \textit{Protein Sci.} \textbf{4}, 561 (1995);
J.N. Onuchic, Z. Luthey-Schulten, and P.G. Wolynes, \textit{Annu. Rev. Phys.
Chem.} \textbf{48}, 545 (1997);
T. Garel, H. Orland, and E. Pitard, \textit{J. Phys. I} \textbf{7},
1201 (1997);
E.I. Shakhnovich, \textit{Curr. Op. Struct. Biol.} \textbf{7}, 29 (1997).
%
\bibitem{tino99}
I.  Tinoco Jr and C. Bustamante, \textit{J. Mol. Biol.} \textbf{293}
271 (1999) and references therein.
%
\bibitem{mcca90}
J.S. McCaskill, \textit{Biopolymers} \textbf{29}, 1105 (1990).
%
\bibitem{rnaalg}
M. Zuker and P. Stiegler, \textit{Nucl. Acid. Res.} \textbf{9},
133 (1981);
I.L. Hofacker, W. Fontana, P.F. Stadler, S. Bonhoeffer, M. Tacker, and
P. Schuster, \textit{Monatshefte f. Chemie} \textbf{125}, 167 (1994).
%
\bibitem{frei86}
S.M. Freier, R. Kierzek, J.A. Jaeger, N. Sugimoto, M.H. Caruthers,
T. Neilson, and D.H. Turner, \textit{Proc. Natl. Acad. Sci. USA} \textbf{83},
9373 (1986).
%
\bibitem{higg96}
P.G. Higgs, \textit{Phys. Rev. Lett.} \textbf{76}, 704 (1996).
%
\bibitem{pagnani}
A. Pagnani, G. Parisi, and E. Ricci-Tersenghi, \textit{Phys. Rev. Lett.}
\textbf{84}, 2026 (2000), \textit{Phys. Rev. Lett.}
\textbf{86}, 1383 (2001); A.K. Hartmann, \textit{ibid.}
\textbf{86}, 1382 (2001).
%
\bibitem{bund99}
R. Bundschuh and T. Hwa, \textit{Phys. Rev. Lett.} \textbf{83}, 1479
(1999).
%
\bibitem{higg00}
P.G. Higgs, \textit{Quart. Rev. Biophys.} \textbf{33},
199 (2000).
%
\bibitem{dege68}
P.G.~de Gennes, \textit{Biopolymers} \textbf{6}, 715 (1968).
%
\bibitem{wate78}
M.S. Waterman, \textit{Advan. in Math. Suppl. Studies} \textbf{1},
167, Academic Press, New York, 1978.
%
\bibitem{bund01}
R. Bundschuh and T. Hwa, in preparation.
%
\bibitem{arra90}
R. Arratia, L. Gordon, and M. Waterman, \textit{Ann. Stat.} \textbf{18},
539 (1990).
%
\bibitem{tang00}
L.-H. Tang and H. Chat\'e, \textit{Phys. Rev. Lett.} \textbf{38},
830 (2001).
%
\end{thebibliography}
\end{document}